# Design and Implementation of Audio Communication System for Social-Humanoid Robot Lumen as an Exhibition Guide in Electrical Engineering Days 2015

Putri Nhirun Rikasofiadewi[1], Ary Setijadi Prihatmanto[2]

*School of Electrical Engineering and Informatic*

*Bandung Institute of Technology, Ganesha Street 10, Bandung 40132, Indonesia*

[1]`putri.nhirun@s.itb.ac.id`
[2]`asetijadi@lskk.ee.itb.ac.id`

*Abstract*— Social Robot Lumen is a humanoid robot created to act like human and be human friend. In this study, Lumen scenario is limited on Lumen as an exhibition guide in Electrical Engineering Days 2015, a seminar and exhibition of electrical engineering undergraduate and graduate student of Bandung Institute of Technology. To be an exhibition guide, Lumen is equipped by Nao robot, a server, and processing applications. Audio communication system is one of the processing applications. The purpose of the system is to create verbal communication that allow Lumen to receive human voice and respond naturally to it. To be able to communicate like a human, audio communication system is built with speech recognition module to transform speech data into text, speech synthesizer module to transform text data into speech, and gender identification module to distinguish adult female and male voice. Speech recognition module is implemented using Google Speech Recognition API, speech synthesizer module is implemented using Acapela engine, and gender identification module implemented by utilizing speech signal feature that is extracted using Fast Fourier Transform algorithm. Hardware used for implementation are Nao robot, computer, and wireless modem.

## I. Introduction

Robots have been developed to have diverse functionality. Not only to supply industrial needs, but robots also developed to interact with human being as a social robot. Hegel et al. require a social robot to have social function and social interface [1]. With this capability of human-robot social interaction, robot is programmed to be a robot guide as well. Lumen is a social-humanoid robot that act and think like human and interact with human.

On this paper, the development will focus on design and implementation of audio communication system to be implemented for Lumen as an exhibition guide in Electrical Engineering Days 2015, a seminar and exhibition of electrical engineering undergraduate and graduate student of Bandung Institute of Technology. With this application, Lumen will be able to hear human voice, explain the product, and answer human questions in the exhibition.

## II. Literature Study

### A. Nao Robot

Nao is humanoid robot from Aldebaran Robotics, a French robotic company. Nao has been used for companion, assistant, and research robot platform. Nao is 573 mm robot. His appearance makes him suitable used with kids. Nao has two stereo loudspeakers used for text-to-speech synthesize and placed as right and left ear. Both its loudspeaker and microphone work with 20-20000 Hz frequency [2]. For synthesizing audio output, Nao uses Acapela text-to-speech (TTS) engine from Acapela group. With this engine, Nao can speak nine languages: English, French, Spanish, German, Italian, Chinese, Japanese, Korean, and Portuguese. Nao uses ALTextToSpeech module inside Nao to make the robot speak. The module sends commands to TTS engine and sent the TTS result to NAO's loudspeaker [3].

### B. Google Speech Recognition

Google has a handful speech recognition program. Their data comes from large number of voice recordings and written search queries uses to predict the words people probably saying [8]. Google use two HTTP connections: one is request for uploading the speech and second is request for access the result. Google's data centre will apply statistical modelling to determine what people saying.

When speak to Google speech recognition, people should speak in normal tempo (the tempo used in daily speaking), people can also talk in accent. Google uses large amount of words to train its recognition software. Google developed its speech recognition using large n-gram language models. With this mode, they can reduce word error rate (WER) between 6% and 10% relative [9].

### C. Speech Signal Feature Extraction

Speech signal carry the information regarding the particular speaker, such as speaker gender. Recognizable frequencies of human voice are between 30 Hz – 3400 Hz with the normal range are between 500 Hz – 2000 Hz [10]. Meanwhile, fundamental frequency of female voice are between 137 Hz – 634 Hz and male voice are between 77 Hz to 482 Hz [11].

Ali et al. studied that frequency in maximum power spectrum is the best feature to classify human gender from speech analysis [4]. In this study, we did not use power spectrum to achieve the frequency. Instead, we use the feature that extracted using Fast Fourier Transform (FFT) algorithm. The FFT result has two variables, voltage in y-axis and frequency in x-axis. The result of FFT is linier with power spectrum. To get power, we will need formula P = v²/R. With P is power, v is voltage, and R is resistance. If R is normalized to 1, we get $P \approx v^2$. This can be illustrated in Fig. 1 and Fig. 2. The frequency in maximum magnitude for both voltage and power is in f3. From this, we can assume FFT result alone is enough without calculating power spectrum.

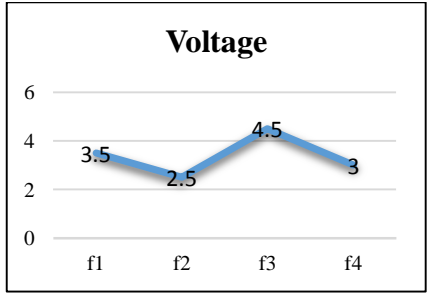

Fig. 1 Illustration of FFT result

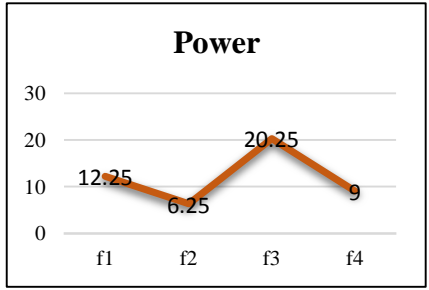

Fig. 2 Illustration of Power Spectrum

FFT is an algorithm for computing the Discrete Fourier Transform (DFT) of a sampled digital signal [5]. One problem calculating DFT is that it is too slow when computed in computer. The FFT solved this problem. The FFT can be accomplished through the use of the Cooley-Tukey algorithm, as in

$$X_k = \sum_{n=0}^{\frac{N}{2}-1} x_{2n} e^{-\frac{2\pi i}{N/2}nk} + e^{-\frac{2\pi i}{N}k} \sum_{n=0}^{\frac{N}{2}-1} x_{2n+1} e^{-\frac{2\pi i}{N/2}nk} \quad (1)$$

In (1), K is integer number. It represents a frequency, while N represents the number of samples, and n is a time-domain sample [6].

III. SYSTEM DESIGN

A. *Lumen Software Architecture*

Lumen is a project developed by writer and team. We created framework described by a Lumen software architecture as in Fig. 3 below.

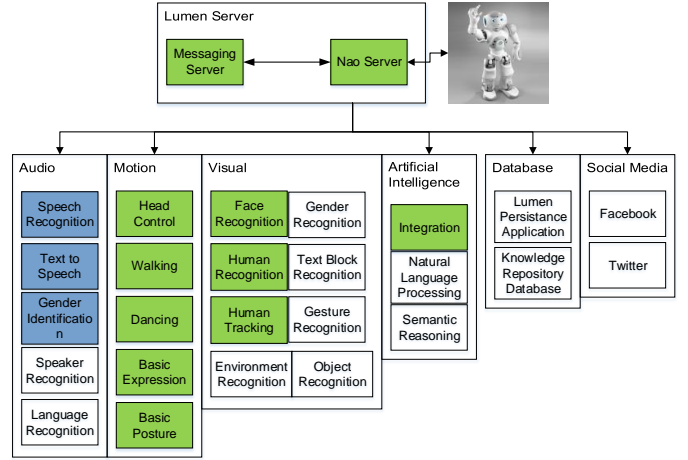

Fig. 3 Software architecture of Lumen

Lumen has seven modules: Lumen server, audio, motion, visual, artificial intelligence, database, and social media [7]. Each module is divided into submodules. Audio, motion, visual, artificial intelligence, database, and social media modules are processing applications.

Lumen server module control communication route between each module to exchange data. Lumen server module also functions as a bridge between each module and Nao robot. The server will get data from each sensor in Nao robot and send commands to the robot. From these functions, Lumen server is divided into two submodules, one is messaging server and second is Nao server.

Visual module functions to do detection and recognition towards human and object. From seven submodules, three are already working: face detection and recognition, human recognition, and human tracking. Artificial intelligence module consists of three submodules and one module, integration, is already working. The integration submodule that use other modules (audio, visual, and others) to create an intelligent. This submodule control when is one module will be called. It also control conversation process in Lumen.

Motion module has five tasks: head control, walking, dancing, basic expression, and basic posture. Head control functions to set head joint related when face tracking submodule on visual module is activated. Lumen can walk forward and backward with its joint set and can dance only with move its right and left hand upward and downward alternately. There are six types of basic postures Lumen can do, shown in Table 1. With basic expression, Lumen can wave its hand to the left and right alternately three times.

Table 1 Lumen basic postures

| No | Postures | Description |
|----|----------|-------------|
| 1  | Stand    | Stand perfectly. |

| 2 | StandInit | Stand with bend its foot a little. |
| 3 | StandZero | Stand with its joint angle is in 0 degree. |
| 4 | Sit | Sit normally. |
| 5 | SitRelax | Sit with its hands lean backwards. |
| 6 | Crouch | Lumen is in crouching position. |

Database and social media modules have not been writer and team focus, yet. With social media modules integrated in the system, Lumen can post any media data to Facebook and Twitter, such as a picture, song, and status, and interact with online. Database module is a place to save any data captured and will integrated to online encyclopedia, such as Wikipedia to do crawling information.

This study only focuses with the blue part that is an audio module. Three submodules that already working on are speech recognition, speech synthesizer or text-to-speech, and gender identification.

### B. Audio Communication System Design

Audio communication system is an audio module allow Lumen to interact with human close to human interaction. With this module, Lumen receive speech signal, send it to speech recognition and gender identification application. Speech recognition submodule will process the speech signal to be text data and gender identification submodule will decide whether the person speaking is female or male. Speech recognition submodule will process the speech signal become text data. This data will be used for artificial intelligence module to decide which command to be sent to motion module or TTS submodule. TTS module will receive the command in text format and process the information to become speech signal in waveform audio file format (WAV). Finally, this audio will be sent to Lumen speaker. That is how Lumen speaks.

We use Google Speech Recognition API to do speech recognition. To be able to use the API, one need internet connection and API key that can be obtained by register to Google. Google Speech Recognition will process audio in Free Lossless Audio Codec (FLAC). Before sent to Google server, audio file must be convert into FLAC format. These are how speech recognition works.

1. Recording a voice with microphone inside Nao robot occur and saved in WAV occurs.
2. This WAV file will be sent to Lumen server.
3. Server program will send the WAV file to speech recognition program to be converted to FLAC format.
4. After converted, the FLAC file sent to Google server.
5. Google server do a recognition and convert the result to a text form.
6. Program request to access the result from Google.
7. Speech recognition program then sent the result to Lumen server.
8. The text can be used for motion command or speak command.

We use Acapela engine to do speech synthesizer. This engine is activated by call ALTextToSpeech module from Nao. Lumen speaks with these process.

1. Lumen server send text data to TTS program.
2. In TTS program, ALTextToSpeech module is called.
3. The module will convert text data into WAV file.
4. The WAV file sent to Lumen server.
5. Lumen server then will send the WAV file to Lumen speakers.

Gender identification application utilize speech signal feature for identifying a person gender. This feature extracted as in described in Fig. 4.

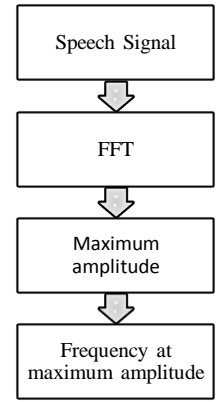

Fig. 4 The sequence operation in extracting speech signal features

Speech signal in is recorded and saved in WAV format. The FFT given to audio file. From this process, we will have a magnitude in x-axis and frequency in y-axis. The next step is to find the maximum magnitude and the frequency at its value.

We use this value to train and recognize human voice. The first stage is training. We collect five samples of female voices and male voice. Data collected are given the FFT to get its threshold value. The value is stored and will be used in the next stage. The second stage is recognition. The voice that incoming will compared to threshold value. If the frequency of recorded voice is higher than threshold value, the program will result "female" and otherwise "male". The more training data are, the better threshold value is, and gender identification will be more accurate. The gender identification block diagram shown in Fig. 5.

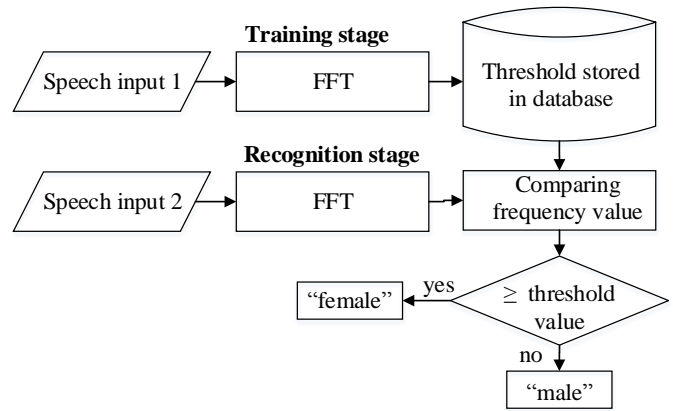

Fig. 5 Block diagram of gender identification system

### C. Conversation Process

Lumen conversation system manage the how the conversation steps are. It represents by a state machine to determine the state of ongoing talks. This state machine has 15 states numbered one until fifteen and each state has its own condition. The state diagram shown in Fig. 6.

Firstly Lumen is in rest position. If there is a person coming and stand in Lumen range view, Lumen will stand and introduce itself. If Lumen recognize detected face, Lumen will greet the people and ask for condition with that person's name. For example, Lumen will say, "good morning, how are you today Putri?" After that, Lumen will wait for the answer. If the person answer "I am fine," Lumen will respond with "I am happy to hear that," but if the person answer "I am not fine," Lumen will respond, "I am sorry to hear that, get well soon." Lumen identify people's gender by listening to their voice. After that, Lumen will ask and address those people's gender. For example, if the person detected is a man, Lumen will ask, "what can I help you, Sir?"

There are several options of answer. First, if the person answer "can you explain about yourself?" Lumen will explain its self and then ask the visitor if they have any questions. Second, if the visitor ask "can we take the picture with you?" Lumen will set a pose to take picture. Third, if the visitor ask "can you dance?" Lumen will dance. Fourth, if the visitor ask "can you sing a song?" Lumen will play a song, and if the visitor say "no thanks", Lumen will wave its hands and sit. If the person face is not in Lumen range view after 20 seconds, Lumen will automatically sit. Another several questions and answers list shown in Table 2.

Table 2 List of questions and answer in Lumen conversation system

| Words contained in the questions | Answers |
|---|---|
| "name" | My name is Lumen |
| "toilet" \| "pray" \| "room" \| "door" | Oh, I'm sorry. I don't know where it is. Maybe you can ask the crew or security. |
| "explain" | My name is Lumen. I am robot guide and you are now in Lumen Super Intelligence Agent stand. I was made to be a tour guide robot. I am able to explain about my stand. I can also amuse you with dancing and singing. I was made by Syarif, Taki, and Putri. That is all about me. |
| "dance" \| "dancing" | Of course I can dance. I will dance a Gangnam Style. Watch carefully, ok. |
| "sing" \| "singing" | Of course I can sing. I will sing Manuk Jajali Song. I will switch my voice to female voice. |
| "exhibition" \| "event" | You are now in Electrical Engineering Days Exhibition. It's an exhibition to show final project of the students. It is held by electrical engineering department of ITB. There are 49 stands of bachelor students including this stand. That is all about EE Days. |
| "made" \| "create" | I am nao robot platform. I am from Aldebaran Robotics, a French robotics company. But, Lumen is programmed by Syarif, Taki, and Putri. |
| "what" \| "do" | I can recognize human face. I can understand human language and respond to them. I can also amuse people with my dancing and singing. I can even walk, you know? |
| "can" \| "walk" | Well, actually I can. But today is a busy day. I need to be in this position for a while. I'm sorry. |
| "can" \| "sit" | Well, actually I can. But today is a busy day. I need to be in this position for a while. I'm sorry. |
| "can" \| "run" | I want to, but no. I can't run. |
| "speak" \| "slow" | I am sorry. I can only talk at this tempo. |
| "tall" \| "height" | I'm about 57 cm high. |
| "weight" \| "fat" | My weight is 5.2 kg. I'm not fat, right? |
| "play" | Well, I want to play with you. But, I can't play around. I need to be in this stand. But I can show you my dancing and singing. |
| "programmed" \| "program" | I was programmed by Syarif, Taki, and Putri. |
| "Lumen" | Lumen is a humanoid robot designed to be an exhibition guide. |
| "Aldebaran" | Aldebaran is a robotic company from French. That's all I can tell you. |
| "old" | I am very young. |
| "weather" | I think it is nice. I don't care anyway. |
| "kind" \| "stand" | There are 49 stands. In each stand presented the final product of electrical engineering students. For more information, you can ask the stand directly. |

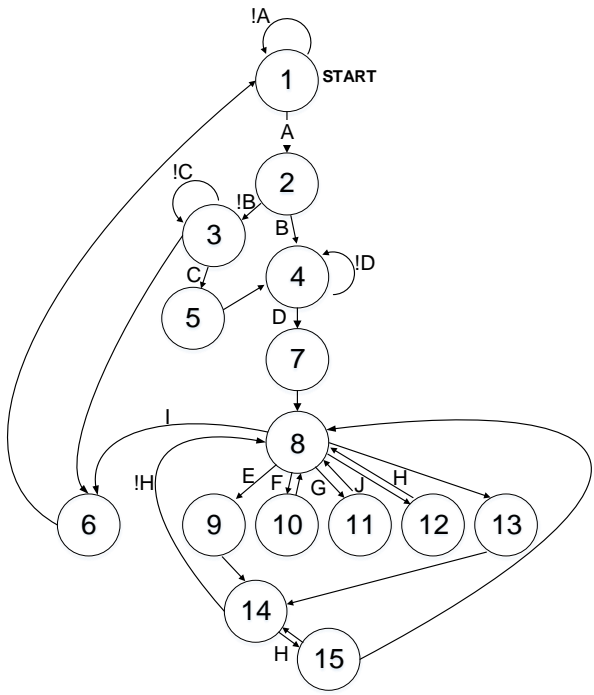

Fig. 6 State machine of Lumen conversation system

Table 3 State and condition explanation of conversation process

| | State | | Condition |
|---|---|---|---|
| 1 | Standby. | A | There is people. |
| 2 | Stand, face recognizing, face tracking, introduction. | !A | No people. |
| 3 | Asking name. | B | Face recognized. |
| 4 | Greeting with people's name | !B | Face unrecognized. |
| 5 | Saving new name and face. | C | There is name respond. |
| 6 | Saying goodbye, waving hand. | !C | No name respond. |
| 7 | Replying greeting. | D | There is greeting respond. |
| 8 | Asking if there's any request. | !D | No greeting respond. |
| 9 | Explaining about the product in stand. | E | Request for explaining product. |
| 10 | Dancing | F | Request for dancing. |
| 11 | Singing. | G | Request for playing music. |
| 12 | Making pose to take picture. | H | Question recognized. |
| 13 | Answer first question. | !H | Question unrecognized. |
| 14 | Asking if there's anything else. | I | User don't respond anything else. |
| 15 | Answering a question. | J | Request for take a picture with Lumen. |

IV. IMPLEMENTATION

All of audio system is implemented in the dotnet framework (.NET) using Visual Studio software with C# programming language.

A. Speech Recognition

Speech recognition application needs stable internet connection because it will be connected to Google server. The recording is done in both silent environment and noisy environment. The result is the application can still recognize the incoming speech signal well. We recorded "hello my friend" three times and "hello" two times both in silent environment and noisy environment. Both yields same result shown in Fig. 7.

```
language : en-us
recognized words : hello my friend

starting recognition...
language : en-us
recognized words : hello my friend

starting recognition...
language : en-us
recognized words : hello my friend

starting recognition...
language : en-us
recognized words : hello

starting recognition...
language : en-us
recognized words : hello
```

Fig. 7 Speech recognition result

One of its flaw is the speech recognition application from Google cannot recognize the words it is too short. For example, if we say "no more", the program will return "a war". We should add some words before or after the sentences, such as "no more, it is enough" to make the program recognize accurately.

B. Speech synthesizer

Speech synthesizer or TTS implemented using Acapela which is an engine inside Nao robot to synthesize a voice. Lumen talk in moderate tempo, but some words are difficult to hear because it's not pronounced well. For example, Lumen pronounce "that" become "de".

C. Gender Identification

We have taken the sampling rate of speech signal as 44 kHz, 16 bit mono. We record speech signal using microphone inside robot with API from Nao robot itself. The recording is done in a room, but not limited to silent condition because in practical Lumen must be able to receive the speech signal in exhibition, which is likely to be noisy environment. The recorded sound was store in WAV sound format. We did not use the header, but we just used the content of the wave file.

Gender identification system operates in two modes: training and recognition mode. We recorded ten samples, with five samples for each male and female. The samples range in 19-24 years old. In the training mode, the speaker asked to say, "Hello my friend, I'm very happy today." The threshold value acquired from the training phase is 598 Hz. The result of gender identification application shown in Table 4 and Table 5 below.

Table 4 Frequency at maximum magnitude of some male and female corresponding English words "Hello my friend, I'm very happy today".

| Speaker ID | Frequency at maximum magnitude (male voice) in Hz | Frequency at maximum magnitude (female voice) in Hz |
|---|---|---|
| 1 | 512 | 623 |
| 2 | 698 | 676 |
| 3 | 497 | 628 |
| 4 | 506 | 576 |
| 5 | 628 | 639 |

Table 5 Detail of recognition result.

| No. of speaker | Speaker gender | Recognize percentage (%) |
|---|---|---|
| 1 | Male | 100 |
| 2 | Male | 0 |
| 3 | Male | 100 |
| 4 | Male | 100 |
| 5 | Male | 100 |
| 6 | Female | 100 |
| 7 | Female | 100 |
| 8 | Female | 100 |
| 9 | Female | 0 |
| 10 | Female | 100 |

The percentage of accuracy rate of recognition has been calculated using (2).

$$\text{Recognition Accuracy (\%)} = \frac{\text{No. of accurately recognized gender}}{\text{No. of correct gender expected}} \times 100 \qquad (2)$$

From the recognition result, number of accurately gender eight samples, while number of correct gender expected is ten samples. The recognition accuracy is = 8*100/10 = 80%.

## V. CONCLUSIONS

1. Speech recognition can be implemented using Google speech recognition API. To get good recognition result, words spoken should not too short.
2. Lumen conversation process is state machine that has state and condition. The more complex state machine in Lumen conversation system, the closer Lumen with the way human interact.
3. The average recognition accuracy of gender identification is 80%. It can be improved with increasing the number of samples.

### ACKNOWLEDGMENT

This development is funded by Indonesia Ministry of National Education and Culture with support from Laboratory of Control System and Computer Bandung Institute of Technology.

### REFERENCES

[1] Hegel, F. et. Al., *Understanding Social Robots,* 2009. https://aiweb.techfak.uni-bielefeld.de/files/2009%20hegel%20ACHI.pdf, [25 April 2015, 14.29 WIB].
[2] Aldebaran Robotics, "*Nao H25 Humanoid Robot Platform."* Des. 2011.
[3] *Aldebaran Documentation* [Online]. Available: http://doc.aldebaran.com/2-1/naoqi/audio/altexttospeech.html#altexttospeech [April 24, 2015].
[4] Ali, Md. Sadek, and Md.Shariful Islam. 2012, "Gender Recognition System Using Speech Signal", *International Journal of Computer Science, Engineering and Information Technology (IJCSEIT)*.
[5] James W. Cooley and John W. Tukey. "An Algorithm for the Machine Calculation of Complex Fourier Series." *Mathematics of Computation*, vol. 19, pp. 297-301, 1965.
[6] C. Oakley. (2011, Nov.) "Development of a Fourier Transform in C#," unpublished [Online]. Available: http://www.egr.msu.edu/classes/ece480/capstone/fall11/group06/style/Application_Note_ChrisOakley.pdf.
[7] A. Syarif, P. Nhirun, and S. Sholata, *Pengembangan Lumen Sebagai Robot Pemandu Pameran, Studi Kasus Electrical Engineering Days 2015,* B400 Engineering Documents, Bandung Institute of Technology, 2015.
[8] *Nextbigfuture, "Google Has Developed Speech-Recognition Technology that Actually Works"* [Online]. Available: http://nextbigfuture.com/2011/04/google-has-developed-speech-recognition.html [April 26, 2015].
[9] C. Chelba, D. Bikel, M. Shugrina, P. Nguyen, and S. Kumar, "Large Scale Language Modeling in Automatic Speech Recognition," Google, 2012.
[10] D. Marshall. (2001). *Human Hearing and Voice* [Online]. Available: https://www.cs.cf.ac.uk/Dave/Multimedia/node271.html
[11] Stemple, J. C., Glaze, L. E., Gerdeman-Klaben, B., *Clinical Voice Pathology, Theory and Management*, 3rd Ed., Canada: Singular Publishing Group, 2000.

# Perancangan dan Implementasi Sistem Komunikasi Audio untuk Lumen Robot Sosial-Humanoid Sebagai Pemandu Pameran Pada *Electrical Engineering Days* 2015

Putri Nhirun Rikasofiadewi[1], Ary Setijadi Prihatmanto[2]

*Sekolah Teknik Elektro dan Informatika*
*Institut Teknologi Bandung, Jalan Ganesha No. 10, Bandung 40132, Indonesia*
[1]`putri.nhirun@s.itb.ac.id`
[2]`asetijadi@lskk.ee.itb.ac.id`

*Abstract*— **Lumen Robot Sosial Robot merupakan robot humanoid yang diciptakan agar dapat bersikap seperti manusia dan menjadi teman bagi manusia. Pada studi ini, Lumen dirancang untuk menjadi pemandu pameran pada *Electrical Engineering Days* 2015, sebuah seminar dan pameran tugas akhir mahasiswa sarjana dan pascasarjana teknik elektro Institut Teknologi Bandung. Untuk dapat menjadi pemandu pameran, Lumen didukung oleh komponen robot Nao, sebuah server, dan beberapa aplikasi pengolah. Sistem komunikasi audio merupakan salah satu aplikasi pengolah yang bertujuan agar Lumen dapat menerima suara manusia dan meresponnya dengan natural, yaitu seperti cara manusia merespon manusia lainnya. Untuk dapat berkomunikasi seperti manusia, sistem komunikasi audio dilengkapi dengan tiga buah modul: *speech recognition* untuk mengubah data suara menjadi teks, *speech synthesizer* untuk mengubah data teks menjadi suara, dan *gender identification* untuk membedakan suara wanita dan pria. *Speech recognition* diimplementasikan menggunakan API Google *Speech Recognition*, *speech synthesizer* diimplementasikan menggunakan *Acapela Engine*, dan modul *gender identification* diimplementasikan menggunakan algoritma *Fast Fourier Transform*. Perangkat keras yang digunakan dalam implementasi Lumen adalah robot Nao, komputer, dan modem *wireless*.**

*Kata kunci*— *robot, audio, sistem komunikasi, speech recognition, speech synthesizer, gender identification, Fast Fourier Transform.*

## I. Pendahuluan

Robot telah dikembangkan untuk memiliki berbagai fungsi. Tidak hanya untuk kebutuhan industri, tetapi juga dikembangkan untuk dapat berinteraksi dengan manusia sebagai robot sosial. Hegel et al. menyatakan bahwa robot dapat dikatakan sebagai robot sosial jika memiliki fungsi dan tampilan sosial [1]. Dengan kemampuan ini, robot dapat pula diprogram untuk menjadi robot pemandu. Lumen merupakan robot a sosial-humanoid yang dapat bertindak dan berpikir seperti manusia dan dapat pula berinteraksi dengan manusia.

Pada makalah ini, akan dijelaskan pengembangan Lumen yang berfokus pada perancangan dan implementasi sistem komunikasi audio untuk Lumen sebagai robot pemandu pada *Electrical Engineering Days* 2015, sebuah seminar dan pameran karya mahasiswa sarjana dan pascasarjana teknik elektro Institut Teknologi Bandung. Dengan aplikasi ini, Lumen dapat mendengar suara manusia, menjelaskan produk di stan Lumen, dan menjawab pertanyaan pengunjung di pameran tersebut.

## II. Studi Literatur

### A. Robot Nao

Nao adalah robot humanoid dari Aldebaran Robotics, sebuah perusahaan robot dari Perancis. Nao telah banyak digunakan sebagai teman, asisten, dan juga penelitian. Tinggi Nao adalah 573 mm. Penampilannya yang kecil dan lucu membuat Nao cocok untuk anak-anak. Nao memiliki dua buah pengeras suara stereo yang digunakan untuk sintesis *text-to-speech* dan terletak di telinga kanan dan kiri robot Nao. Mikrofon dan pengeras suara Nao bekerja pada rentang frekuensi 20-20000 Hz [2]. Untuk mensintesis suara, Nao menggunakan mesin *text-to-speech* (TTS) Acapela dari Acapela Group. Dengan mesin ini, Nao dapat berbicara sembilan Bahasa, yaitu bahasa Inggris, Perancis, Spanyol, German, Italia, Cina, Jepang, Korea, dan Portugis. Nao menggunakan modul ALTextToSpeech yang untuk dapat berbicara. Modul tersebut akan mengirimkan perintah ke mesin TTS dan mengirimkan hasilnya ke pengeras suara Nao [3].

### B. Google Speech Recognition

Google memiliki program *speech recognition*. Data yang dimiliki Google berasal dari sejumlah besar rekaman suara dan pencarian *query* tertulis untuk memprediksi kata-kata yang mungkin diucapkan oleh orang-orang [8]. Google menggunakan dua buah koneksi HTTP. Koneksi pertama adalah permintaan untuk mengunduh sinyal suara ke server Google dan kedua adalah untuk mengakses hasil rekognisi tersebut. Kemudian, pusat data Google akan menggunakan pemodelan statistik tertentu untuk menentukan isi perkataan yang diucapkan.

Ketika merekam suara untuk Google *speech recognition*, suara yang diucapkan harus dalam tempo normal (tempo yang biasa diucapkan pada percakapan sehari-hari). Google *speech recognition* juga dapat mengenali bahasa yang diucapkan dengan aksen. Google menggunakan sejumlah besar data kata-kata untuk melatih program rekognisi yang mereka miliki. Google mengembangkan *speech recognition* menggunakan model bahasa *large n-gram*. Dengan model ini, mereka dapat mengurangi jumlah kata eror atau *word error rate* (WER) hingga relatif 6% sampai 10% [9].

*C. Ekstraksi Fitur Sinyal Suara*

Sinyal suara membawa informasi terkait orang yang berbicara, seperti jenis kelamin pembicara. Frekuensi manusia yang dapat dikenali berada pada rentang 30 Hz – 3400 Hz dengan rentang normal berada pada 500 Hz – 2000 Hz [10]. Sementara itu, frekuensi dasar suara perempuan berada pada rentang 137 Hz – 634 Hz dan untuk laki-laki berada pada rentang 77 Hz - 482 Hz [11].

Ali et al. mempelajari bahwa frekuensi pada spektrum daya maksimum adalah fitur terbaik untuk mengklasifikasikan jenis kelamin manusia dari analisis sinyal suara [4]. Pada pengembangan Lumen ini, penulis tidak menggunakan skpektrum daya untuk mendapatkan frekuensi tersebut, melainkan menggunakan *magnitude* hasil algoritma *Fast Fourier Transform* (FFT). Hasil dari FFT adalah dua buah variabel yang diwakili oleh tegangan pada sumbu-y dan frekuensi pada sumbu-x. Hasil dari FFT linier dengan spektrum daya. Untuk mendapatkan daya, digunakan rumus $P = v^2/R$, dengan P adalah daya, v adalah tegangan, dan R adalah nilai hambatan. Jika nilai R dinormalisasi menjadi 1, akan didapatkan $P \approx v^2$. Hal ini dapat diilustrasikan pada Gambar 1 dan Gambar 2. Pada gambar tersebut, frekuensi pada nilai maksimum, baik untuk tegangan dan daya adalah f3. Dari gambar ini dapat diasumsikan bahwa hasil FFT saja sudah cukup untuk mendapatkan frekuensi pada nilai tegangan maksimum tanpa perlu menghitung spektrum daya karena akan diperoleh nilai frekuensi yang sama.

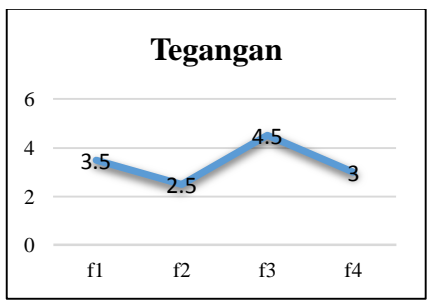

Gambar 1 Ilustrasi hasil perhitungan FFT

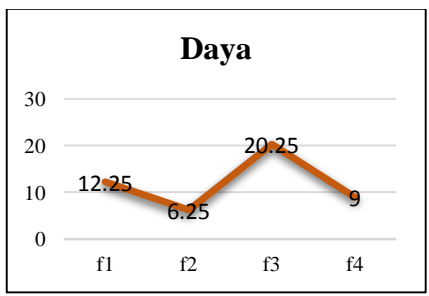

Gambar 2 Ilustrasi hasil perhitungan spektrum daya

FFT merupakan algoritma untuk menghitung *Discrete Fourier Transform* (DFT) dari sinyal digital yang sudah disampel [5]. Salah satu masalah dalam menghitung DFT adalah prosesnya lambat jika dilakukan di komputer. FFT menyelesaikan masalah tersebut. FFT dapat dilakukan dengan menggunakan algorima Cooley-Tukey, seperti pada (1).

$$X_k = \sum_{n=0}^{\frac{N}{2}-1} x_{2n} e^{-\frac{2\pi i}{N/2}nk} + e^{-\frac{2\pi i}{N}k} \sum_{n=0}^{\frac{N}{2}-1} x_{2n+1} e^{-\frac{2\pi i}{N/2}nk} \quad (1)$$

Pada (1), K adalah bilangan integer yang merepresentasikan frekuensi, sementara N merepresentasikan jumlah sampel, dan n merupakan sampel dalam domain waktu [6].

### III. PERANCANGAN SISTEM

*A. Arsitektur Perangkat Lunak Lumen*

Lumen merupakan sebuah proyek yang dikembankan oleh penulis dan tim. Kami membuat sebuah kerangka kerja yang digambarkan oleh arsitektur perangkat lunak Lumen seperti pada Gambar 3.

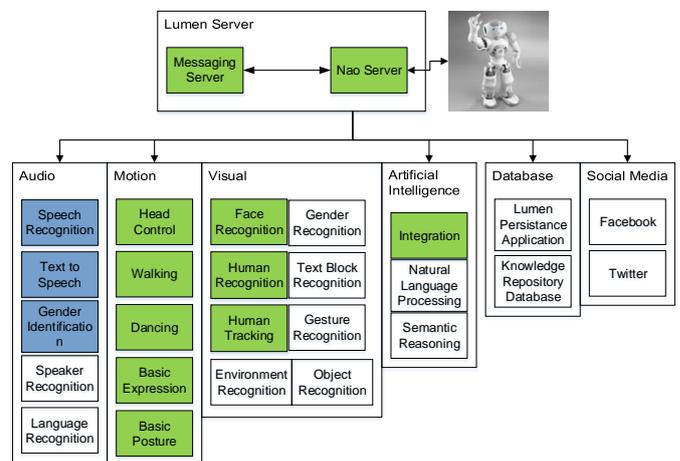

Gambar 3 Arsitektur perangkat lunak Lumen

Lumen memiliki tujuh modul, yaitu Lumen *server,* audio, *motion*, visual, *artificial intelligence*, *database*, and *social media* [7]. Setiap modul memiliki submodul masing-masing. Modul audio, *motion*, visual, *artificial intelligence, database*, dan *social media* merupakan aplikasi pemroses.

Modul Lumen *server* mengatur rute komunikasi antara setiap modul untuk pertukaran data. Modul ini juga berfungsi sebagai jembatan antara tiap modul dengan robot Nao. *Server* akan mengambil data dari tiap sensor pada robot Nao dan mengirim perintah ke robot. Dari fungsi tersebut, Lumen *server* dibagi menjadi dua submodul, pertama adalah *messaging server* dan kedua adalah Nao *server.*

Modul visual berfungsi untuk mendeteksi dan mengenali manusia dan benda. Dari tujuh submodul pada modul visual, tiga submodul telah berhasil diimplementasikan, yaitu deteksi dan pengenalan wajah, pengenalan manusia, dan pelacakan muka manusia. Modul *artificial intelligence* terdiri atas tiga submodul, satu diantaranya sudah berjalan, yaitu *integration*.

Submodul ini menggunakan modul-modul lainnya (audio, visual, dan lain-lain) untuk membuat kecerdasan. Submodul ini akan mengatur kapan modul lainnya akan dipanggil. Submodul ini juga mengatur proses percakapan pada Lumen.

Modul *motion* memiliki lima tugas, yaitu *head control, walking, dancing, basic expression,* dan *basic posture. Head control* berfungsi untuk mengatur sendi kepala yang terkait dengan submodule *face tracking* pada saat diaktifkan. Lumen dapat berjalan ke depan dan ke belakang dengan cara mengatur sudut sendi-sendi dan dapat menari dengan kedua tangannya saja. Terdapat enam tipe *basic postures* yang dimiliki Lumen, seperti yang ditunjukkan oleh Tabel 1. Dengan ekspresi dasar, Lumen dapat melambaikan tangannya ke kiri dan ke kanan bergantian sebanyak tiga kali.

Tabel 1 B*asic postures* Lumen

| No | Postur | Keterangan |
|---|---|---|
| 1 | *Stand* | Berdisi sempurna. |
| 2 | *StandInit* | Berdiri dengan sedikit menekuk kaki. |
| 3 | *StandZero* | Berdiri dengan sudut sendinya berada pada 0 derajat. |
| 4 | *Sit* | Duduk normal. |
| 5 | *SitRelax* | Duduk dengan kedua tangannya bersandar ke belakang. |
| 6 | *Crouch* | Lumen dalam posisi meringkuk. |

Modul *database* dan *social media* tidak menjadi fokus penulis dan tim saat ini. Dengan modul *social media* yang terintegrasi ke dalam sistem, Lumen dapat memasukkan data media apapun ke Facebook and Twitter, seperti gambar, lagu, dan status, dan juga berinteraksi *online*. Modul *database* merupakan tempat untuk menyimpan semua data yang diperoleh oleh robot Nao dan juga terintegrasi dengan kamus *online*, seperti Wikipedia, untuk melakukan pencarian informasi.

Penelitian yang penulis bahas dalam makalah ini hanya berfokus pada bagian berwarna biru, yaitu modul audio. Tiga submodul yang telah berjalan, yaitu *speech recognition, speech synthesizer or text-to-speech,* dan *gender identification*.

### B. Perancangan Sistem Komunikasi Audio

Sistem komunikasi audio merupakan sebuah modul yang membuat Lumen dapat berinteraksi mendekati cara manusia berinteraksi. Dengan modul ini, Lumen menerima sinyal suara, mengirimnya ke aplikasi *speech recognition* dan *gender identification application*. Submodul *speech recognition submodul* akan memproses sinyal suara menjadi data teks dan submodul *gender identification* akan menentukan apakah orang yang berbicara adalah wanita atau pria. Submodul *speech recognition* akan memproses sinyal suara menjadi data teks. Data ini akan digunakan oleh modul *artificial intelligence* untuk memutuskan perintah apa yang akan dikirim ke modul *motion* atau submodul TTS. Modul TTS akan menerima perintah dalam format teks dan memproses informasi tersebut untuk diolah menjadi sinyal suara dalam format file audio *waveform* (WAV). Terakhir, file audio tersbut akan dikirim ke pengeras suara Lumen. Dengan cara inilah Lumen berbicara.

Penulis menggunakan API dari Google *Speech Recognition* untuk melakukan rekognisi sinyal suara. Untuk dapat menggunakan API tersebut, diperlukan adanya koneksi internet dan kunci API yang dapat diperoleh dengan mendaftar ke Google terlebih dahulu. *Google Speech Recognition* akan memproses audio dalam format *Free Lossless Audio Codec* (FLAC). Sebelum dikirim ke server Google, file audio harus terlebih dulu dikonversi ke format FLAC. Berikut ini merupakan cara *speech recognition* berkeja.

1. Perekaman suara dengan mikrofon yang ada di dalam robot Nao terjadi dan suara disimpan dalam WAV.
2. File WAV ini akan dikirim ke server Lumen.
3. Program server akan mengirim file WAV file ke program *speech recognition* untuk dikonversi ke format FLAC.
4. Setelah dikonversi, file FLAC dikirim ke server Google.
5. Server Google melakukan rekognisi dan mengkonversi hasilnya menjadi bentuk teks.
6. Program meminta akses mengambil hasil rekognisi dari Google.
7. Program *speech recognition* mengirim hasilnya ke server Lumen.
8. Data teks tersebut dapat digunakan untuk perintah bergerak atau perintah berbicara.

Penulis menggunakan mesin Acapela untuk mensintesis suara. Mesin ini diaktifkan dengan memanggil modul ALTextToSpeech dari robot Nao. Lumen berbicara dengan proses sebagai berikut.

1. Server Lumen mengirim data teks ke program TTS.
2. Dalam program TTS, modul ALTextToSpeech dipanggil.
3. Modul tersebut akan mengkonversi data teks ke file WAV.
4. File tersebut kemudian dikirim ke server Lumen.
5. Server Lumen akan mengirim file WAV ke pengeras suara Lumen.

Aplikasi *gender identification* memanfaatkan fitur sinyal suara untuk mengidentifikasi jenis kelamin seseorang. Fitur ini diekstraksi seperti yang digambarkan pada Gambar 4.

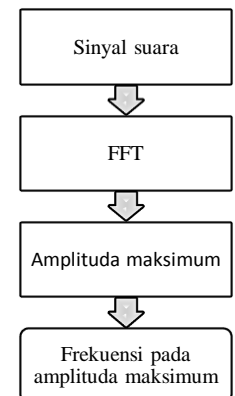

Gambar 4 Urutan operasi dalam ekstraksi fitur sinyal suara

Sinyal suara direkam dan disimpan dalam format WAV. Kemudian, dilakukan proses FFT terhadap file audio yang direkam tersebut. Dari proses ini akan dihasilkan sebuah besaran amplitudo dalam sumbu-x dan frekuensi dalam sumbu-y. Langkah selanjutnya adalah mencari amplitudo terbesar dan frekuensi pada amplitudo tersebut.

Penulis menggunakan nilai ini untuk melatih dan mengenali suara manusia. Tahap pertama adalah pelatihan. Penulis mengumpulkan lima suara wanita dan lima suara pria. Data yang terkumpul diberikan proses FFT untuk mendapatkan nilai *threshold*. Nilai ini disimpan dan akan digunakan pada tahap selanjutnya. Tahap kedua adalah rekognisi atau pengenalan. Suara yang datang akan dibandingkan dengan nilai *threshold*. Jika frekuensi dari suara yang masuk lebih besar daripada nilai *threshold*, program akan menghasilkan "female" sebaliknya "male". Semakin banyak sampel yang digunakan untuk pelatihan, semakin baik nilai *threshold* yang diperoleh, dan semakin akurat identifikasi jenis kelamin tersebut. Diagram blok dari identifikasi jenis kelamin ditunjukkan pada Gambar 5.

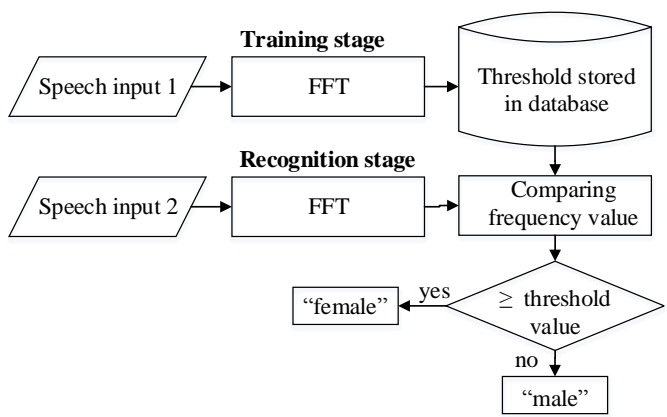

Gambar 5 Diagram blok sistem *gender identification*

### C. Proses Percakapan

Sistem percakapan Lumen mengatur bagaimana tahapan-tahapan dalam melakukan percakapan tersebut. Sistem ini direpresentasikan dengan *state machine*. *State machine* ini memiliki 15 posisi dan setiap posisi memiliki kondisi masing-masing. Diagram dari posisi tersebut ditunjukkan oleh Gambar 6.

Pertama-tama, Lumen berada dalam posisi istirahat. Jika ada orang datang dan berdiri pada jarak penglihatan Lumen, Lumen akan berdiri dan memperkenalkan dirinya. Jika Lumen mengenali wajah orang yang datang tersebut, Lumen akan menyapa orang tersebut dengan namanya. Contohnya, Lumen akan mengatakan, "good morning, how are you today Putri?" Setelah itu, Lumen menunggu jawaban dari orang tersebut. Jika mendengar jawaban "I am fine," Lumen akan merespon dengan "I am happy to hear that," tetapi jika mendengar jawaban "I am not fine," Lumen akan merespon dengan, "I am sorry to hear that, get well soon." Lumen mengidentifikasi jenis kelamin manusia dengan mendengarkan suara yang masuk. Setelah itu, Lumen akan bertanya dan memanggil orang tersebut dengan sapaan berdasarkan jenis kelaminnya. Contohnya adalah jika orang terdeteksi sebagai pria, Lumen akan bertanya "what can I help you, Sir?"

Terdapat beberapa pilihan jawaban yang dikenali oleh Lumen. Pertama, jika orang tesebut menjawab "can you explain about yourself?" Lumen akan menjelaskan tentang dirinya dan kemudian bertanya kepada pengunjung tersebut apakah dia memiliki pertanyaan. Kedua, jika pengunjung bertanya "can we take the picture with you?" Lumen akan memasang pose untuk berfoto. Ketiga, jika pengunjung bertanya "can you dance?" Lumen akan menari. Keempat, jika pengunjung bertanya "can you sing a song?" Lumen akan memutar sebuah lagu, dan jika pengunjung berkata "no thanks", Lumen akan melambaikan tangannya kemudian duduk. Jika wajah pengunjung tidak berada dalam jarak pandang Lumen selama 20 detik, Lumen akan duduk secara otomatis. Beberapa pertanyaan dan jawaban lainnya ditunjukkan oleh Tabel 2.

Tabel 2 Daftar pertanyaan dan jawaban dalam sistem percakapan Lumen

| Kata yang terkandung dalam pertanyaan | Jawaban |
|---|---|
| "name" | My name is Lumen |
| "toilet" \| "pray" \| "room" \| "door" | Oh, I'm sorry. I don't know where it is. Maybe you can ask the crew or security. |
| "explain" | My name is Lumen. I am robot guide and you are now in Lumen Super Intelligence Agent stand. I was made to be a tour guide robot. I am able to explain about my stand. I can also amuse you with dancing and singing. I was made by Syarif, Taki, and Putri. That is all about me. |
| "dance" \| "dancing" | Of course I can dance. I will dance a Gangnam Style. Watch carefully, ok. |
| "sing" \| "singing" | Of course I can sing. I will sing Manuk Dadali Song. I will switch my voice to female voice. |
| "exhibition" \| "event" | You are now in Electrical Engineering Days Exhibition. It's an exhibition to show final project of the students. It is held by electrical engineering department of ITB. There are 49 stands of bachelor students including this stand. That is all about EE Days. |
| "made" \| "create" | I am nao robot platform. I am from Aldebaran Robotics, a French robotics company. But, Lumen is programmed by Syarif, Taki, and Putri. |
| "what" \| "do" | I can recognize human face. I can understand human language and respond to them. I can also amuse people with my dancing and singing. I can even walk, you know? |
| "can" \| "walk" | Well, actually I can. But today is a busy day. I need to be in this position for a while. I'm sorry. |
| "can" \| "sit" | Well, actually I can. But today is a busy day. I need to be in this position for a while. I'm sorry. |
| "can" \| "run" | I want to, but no. I can't run. |
| "speak" \| "slow" | I am sorry. I can only talk at this tempo. |
| "tall" \| "height" | I'm about 57 cm high. |
| "weight" \| "fat" | My weight is 5.2 kg. I'm not fat, right? |
| "play" | Well, I want to play with you. But, I can't play around. I need to be in this stand. But I can show you my dancing and singing. |
| "programmed" \| "program" | I was programmed by Syarif, Taki, and Putri. |
| "Lumen" | Lumen is a humanoid robot designed to be an exhibition guide. |
| "Aldebaran" | Aldebaran is a robotic company from French. That's all I can tell you. |
| "old" | I am very young. |
| "weather" | I think it is nice. I don't care anyway. |
| "kind" \| "stand" | There are 49 stands. In each stand presented the final product of electrical engineering students. For more information, you can ask the stand directly. |

| 2 | Stand, face recognizing, face tracking, introduction. | !A | No people. |
|---|---|---|---|
| 3 | Asking name. | B | Face recognized. |
| 4 | Greeting with people's name | !B | Face unrecognized. |
| 5 | Saving new name and face. | C | There is name respond. |
| 6 | Saying goodbye, waving hand. | !C | No name respond. |
| 7 | Replying greeting. | D | There is greeting respond. |
| 8 | Asking if there's any request. | !D | No greeting respond. |
| 9 | Explaining about the product in stand. | E | Request for explaining product. |
| 10 | Dancing | F | Request for dancing. |
| 11 | Singing. | G | Request for playing music. |
| 12 | Making pose to take picture. | H | Question recognized. |
| 13 | Answer first question. | !H | Question unrecognized. |
| 14 | Asking if there's anything else. | I | User don't respond anything else. |
| 15 | Answering a question. | J | Request for take a picture with Lumen. |

IV. IMPLEMENTASI

Semua sistem audio diimplementasikan dalam kerangka kerja dotnet (.NET) menggunakan perangkat lunak Visual Studio dan dengan bahasa pemrograman C#.

A. Speech Recognition

Aplikasi *speech recognition* membutuhkan koneksi internet yang stabil karena aplikasi ini akan melakukan koneksi ke server Google. Perekaman dilakukan pada lingkungan yang tenang dan berisik. Hasilnya menunjukkan aplikasi ini dapat mengenali sinyal suara yang masuk dengan baik. Penulis merekam kata "hello my friend" sebanyak tiga kali and kata "hello" sebanyak dua kali masing-masing pada lingkungan yang tenang dan berisik. Keduanya menghasilkan rekognisi yang sama, seperti yang ditunjukkan oleh Gambar 7.

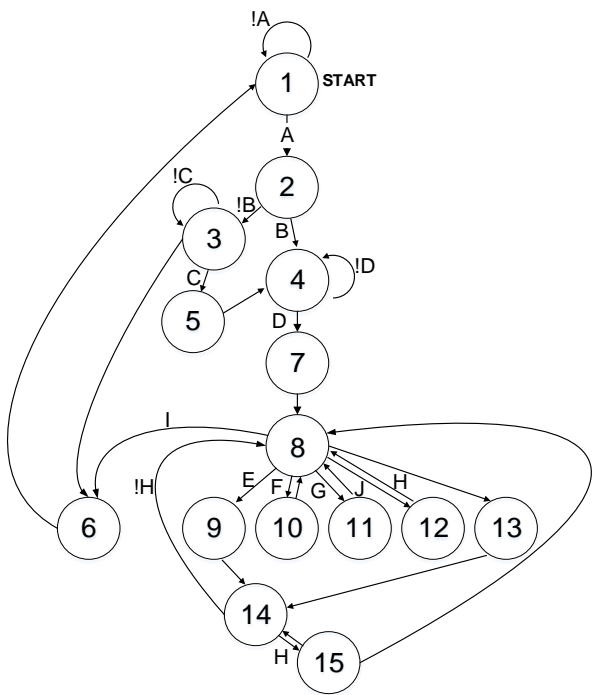

Gambar 7 Hasil aplikasi *speech recognition*

Salah satu kekurangan aplikasi *speech recognition* dari Google adalah tidak dapat mengenali kata jika kata tersebut

Gambar 6 *State machine* dari sistem percakapan Lumen

Tabel 3 Penjelasan posisi dan kondisi dari proses percakapan Lumen

| | Posisi | | Kondisi |
|---|---|---|---|
| 1 | Standby. | A | There is people. |

terlalu pendek. Contohnya, jika penulis mengatakan "no more", program akan mengembalikan "a war". Oleh karena itu, harus ditambahkan beberapa kata sebelum atau sesudah kalimat pendek tersebut, seperti "no more, it is enough" agar program dapat melakukan rekognisi secara akurat.

*B. Speech synthesizer*

Speech synthesizer atau TTS diimplementasikan menggunakan mesin Acapela yang terdapat di dalam robot Nao. Lumen berbicara dalam tempo yang sedang, tetapi beberapa kata-kata terdengar tidak jelas karena tidak terucapkan dengan baik. Contohnya, Lumen mengucapkan "that" menjadi "de".

*C. Gender Identification*

Penulis mengambil sinyal suara dengan frekuensi sampling sebesar 44 kHz, 16 bit mono. Sinyal suara direkam dengan menggunakan mikrofon yang ada di robot Nao. Proses rekaman dilakukan di dalam ruangan, tetapi tidak dibatas pada kondisi tenang karena pada penggunaannya, Lumen harus dapat menerima sinyal suara di pameran yang kemungkinan besar memiliki lingkungan dengan derau tinggi. Suara yang telah direkam disimpan dalam format suara WAV tanpa *header*.

Sistem *gender identification* beroperasi dalam dua modus: pelatihan dan pengenalan. Penulis merekam sepuluh sampel, masing-masing lima orang wanita dan lima orang pria. Umur dari sampel yang diambil berkisar antara 19-24 tahun. Pada modus pelatihan, sampel akan diminta untuk berkata, "Hello my friend, I'm very happy today." Nilai *threshold* yang didapatkan dari modus pelatihan adalah 598 Hz. Hasil dari aplikasi *gender identification* ditunjukkan oleh Tabel 4 dan Tabel 5.

Tabel 4 Frekuensi pada amplituda maksimum dari beberapa sampel laki-laki dan perempuan dengan kalimat "Hello my friend, I'm very happy today".

| Pembicara | Frekuensi pada amplituda maksimum (pria) in Hz | Frekuensi pada amplituda maksimum (wanita) in Hz |
|---|---|---|
| 1 | 512 | 623 |
| 2 | 698 | 676 |
| 3 | 497 | 628 |
| 4 | 506 | 576 |
| 5 | 628 | 639 |

Tabel 5 Detil hasil rekognisi.

| Nomor pembicara | Jenis kelamin pembicara | Persentase dikenali (%) |
|---|---|---|
| 1 | Male | 100 |
| 2 | Male | 0 |
| 3 | Male | 100 |
| 4 | Male | 100 |
| 5 | Male | 100 |
| 6 | Female | 100 |
| 7 | Female | 100 |
| 8 | Female | 100 |
| 9 | Female | 0 |
| 10 | Female | 100 |

Persentase akurasi rekognisi dihitung dengan menggunakan (2).

$$\text{Akurasi pengenalan (\%)} = \frac{\text{jumlah dikenal benar}}{\text{jumlah diharapkan benar}} \times 100 \qquad (2)$$

Dari hasil pengenalan, jumlah yang dikenali sebanyak delapan orang, sedangkan jumlah yang diharapkan benar sebanyak sepuluh orang, sehingga akurasi pengenalan adalah 8*100/10 = 80%.

## V. KESIMPULAN

1. *Speech recognition* dapat diimplementasikan menggunakan API Google Speech Recognition. Untuk mendapatkan hasil pengenalan yang baik, kata yang diucapkan tidak boleh terlalu pendek.
2. Proses percakapan pada Lumen merupakan sebuah *state machine* yang memiliki posisi dan kondisi tertentu. Semakin kompleks *state machine* pada Lumen, semakin dekat Lumen dengan cara manusia berinteraksi.
3. Rata-rata akurasi rekognisi dari *gender identification* sebesar 80%. Nilai ini dapat ditingkatkan dengan menambah jumlah sampel.

ACKNOWLEDGMENT

Pengembangan proyek Lumen ini didanai oleh Menteri Pendidikan dan Kebudayaan Indonesia dengan dukungan dari Laboratorium Sistem Kendali dan Komputer Institut Teknologi Bandung.

REFERENSI

[1] Hegel, F. et. Al., *Understanding Social Robots,* 2009. https://aiweb.techfak.uni-bielefeld.de/files/2009%20hegel%20ACHI.pdf, [25 April 2015, 14.29 WIB].
[2] Aldebaran Robotics, "*Nao H25 Humanoid Robot Platform."* Des. 2011.
[3] *Aldebaran Documentation* [Online]. Tersedia: http://doc.aldebaran.com/2-1/naoqi/audio/altexttospeech.html#altexttospeech [April 24, 2015].
[4] Ali, Md. Sadek, and Md.Shariful Islam. 2012, "Gender Recognition System Using Speech Signal", *International Journal of Computer Science, Engineering and Information Technology (IJCSEIT)*.
[5] James W. Cooley and John W. Tukey. "An Algorithm for the Machine Calculation of Complex Fourier Series." *Mathematics of Computation*, vol. 19, pp. 297-301, 1965.
[6] C. Oakley. (2011, Nov.) "Development of a Fourier Transform in C#," unpublished [Online]. Tersedia: http://www.egr.msu.edu/classes/ece480/capstone/fall11/group06/style/Application_Note_ChrisOakley.pdf.
[7] A. Syarif, P. Nhirun, and S. Sholata, *Pengembangan Lumen Sebagai Robot Pemandu Pameran, Studi Kasus Electrical Engineering Days 2015,* Dokumen B400, Institut Teknologi Bandung, 2015.
[8] *Nextbigfuture, "Google Has Developed Speech-Recognition Technology that Actually Works"* [Online]. Tersedia: http://nextbigfuture.com/2011/04/google-has-developed-speech-recognition.html [April 26, 2015].
[9] C. Chelba, D. Bikel, M. Shugrina, P. Nguyen, and S. Kumar, "Large Scale Language Modeling in Automatic Speech Recognition," Google, 2012.
[10] D. Marshall. (2001). *Human Hearing and Voice* [Online]. Tersedia: https://www.cs.cf.ac.uk/Dave/Multimedia/node271.html
[11] Stemple, J. C., Glaze, L. E., Gerdeman-Klaben, B., *Clinical Voice Pathology, Theory and Management*, 3rd Ed., Canada: Singular Publishing Group, 2000.